\documentclass[twocolumn,english,aps,prl,groupedaddress,superscriptaddress]{revtex4}
\usepackage{graphicx,epsfig,units}
\usepackage{xcolor} 
\usepackage{soul} 
\usepackage{amsmath,amsfonts,mathrsfs,amsbsy,bm,babel}

\begin{document}

\title{Magnetic Order and Competition With Superconductivity in (Er-Ho)Ni$_{2}$B$_{2}$C}

\author{Suleyman Gundogdu}
\affiliation{Physics Engineering Department, Istanbul Technical University, 34469, Maslak, Istanbul, Turkey}

\author{J. Patrick Clancy}
\affiliation{Department of Physics and Astronomy, McMaster University, Hamilton, Ontario L8S 4M1 Canada }

\author{Guangyong Xu}
\affiliation{NIST Center for Neutron Research, National Institute of Standards and Technology, Gaithersburg, Maryland 20899, USA}

\author{Yang Zhao}
\affiliation{NIST Center for Neutron Research, National Institute of Standards and Technology, Gaithersburg, Maryland 20899, USA}

\author{Paul A. Dube}
\affiliation{Brockhouse Institute for Materials Research, Hamilton, ON L8S 4M1, Canada}

\author{Tufan C. Karalar}
\affiliation{Electronics and Communication Engineering Department, Istanbul Technical University, 34469, Maslak, Istanbul, Turkey}

\author{Beong Ki Cho}
\affiliation{Gwangju Institute of Science and Technology, GIST, S. Korea} 

\author{Jeffrey W. Lynn}
\affiliation{NIST Center for Neutron Research, National Institute of Standards and Technology, Gaithersburg, Maryland 20899, USA}

\author{M.~Ramazanoglu}
\affiliation{Physics Engineering Department, Istanbul Technical University, 34469, Maslak, Istanbul, Turkey}
\affiliation{Brockhouse Institute for Materials Research, Hamilton, ON L8S 4M1, Canada}

\begin{abstract}
The rare earth magnetic order in pure and doped Er$_{(1-x)}$Ho$_{x}$Ni$_2$B$_2$C (x~=~0,~0.25,~0.50,~0.75,~1) single crystal samples was investigated using magnetization and neutron diffraction measurements. Superconducting quaternary borocarbides, $R$Ni$_2$B$_2$C where R~=~Er, Ho , are both magnetic intermetallic superconductors with the transition temperatures $\sim$ 10 K. These compounds also develop magnetic order in the vicinity of this temperature. Depending on  the rare earth composition  the coupling between superconductivity and magnetism creates several phases, ranging from a reentrant superconductor with a mixture of commensurate and incommensurate antiferromagnetism to a total incommensurate antiferromagnetic spin modulation with a weak ferromagnetic state. All of these phases coexist with superconductivity. RKKY  magnetic interactions are used to describe the magnetic orders in the pure compounds. However, the doping of Ho on Er (or Er on Ho) sites which have two strong magnetic moments with two different easy directions creates new and complicated magnetic modulations with possible local disorder effects. One fascinating effect is the development of an induced magnetic state resembling the pure and doped $R_2$CuO$_4$, R~=~Nd and Pr.

\end{abstract}

\date{August 11, 2020}
\maketitle

Rare earth ($R$) nickel borocarbides $R$Ni$_2$B$_2$C have managed to stay a current research topic for more than 25 years. The reason for this longtime attraction is related to their magnetic low temperature states which occur in the vicinity of temperatures with superconductivity (SC) \cite{Canfield, Gupta}. There are other magnetic superconductors available for scientists, however, in the quaternary borocarbides depending on choice of $R$ element there is a wide range of magnetism ranging from antiferromagnetic (AFM) short modulations to a ferromagnetic (FM) state \cite{Lynn}. These magnetic states which naturally compete with SC create exotic phenomena when they exhibit long range order simultaneously with the superconducting state. This competition is important when SC, which is a $q$~=~0 ordered magneto-electronic state, interplays with magnetism which usually orders along a $q$~$\neq$~0 vector (except for FM) \cite{kawano}. The availability of high quality single-crystal samples for all $R$ members of the $R$Ni$_2$B$_2$C family makes this system an indispensable quantum-magnetism paradigm to investigate.
The magnetic interactions in $R$Ni$_2$B$_2$C emerge via the Ruderman-Kittel-Kasuya-Yosida (RKKY) model interactions that couple the local spins to the itinerant conduction electrons \cite{Amici}. 
The $R$Ni$_2$B$_2$C crystal structure is body centered tetragonal (I4/mmm at room temperature) for both $R$~=~Er and Ho. Plus, the crystal unit cell dimensions are very close to each other ($a$~$\sim$~3.518 $\AA$ , $c$~$\sim$~10.527 $\AA$ for $R$~=~Ho and   
$a$~$\sim$~3.502 $\AA$ , $c$~$\sim$~10.558 $\AA$ for $R$~=~Er) so that doping on the $R$ site with Er and Ho together is permissible across the entire concentration range \cite{Lynn2}. In the pure $R$~=~Ho case, the Ho moments are coupled ferromagnetically in the $a\textrm{-}b$ plane, forming FM sheets which are antiferromagnetically stacked along the c-axis. This pattern of moments produces antiferromagnetic modulations with q$_c$~=~0.915 $c^*$ that exists over a short temperature interval upon initial ordering and then a commensurate antiferromagnetic (C-AFM) structure which coexists with SC.   
The associated ordering temperatures for the development of the C-AFM and the SC phases coincide at the same values of T~$\sim$~8~K \cite{GoldmanII,Grigereit}.  
In addition, the first single crystal studies revealed an additional incommensurate (IC) a-axis modulation over a narrow temperature range between 5 K $ \leq $  T $ \leq $ 7 K with a wave-vector of q$_a$~=~0.585 $a^*$ \cite{GoldmanII}.
Many additional details of the magnetic structures have been elucidated via high resolution neutron and x-ray investigations on single crystals, including a small structural distortion from tetragonal to orthorhombic settings \cite{Sazonov, Kressig, Hill}.
$Er$Ni$_2$B$_2$C shows a transversely polarized spin density wave modulation which is always incommensurate with a wave-vector of q$_a$~=~0.55 $a^*$ at T$_N$ ~$\sim$~6~K \cite{Choi}. This means the easy direction for Er moments is along the $b$-axis. 
Higher order reflections were observed for this magnetic state reflecting the squaring up of the spin density wave order, along with the breaking of time reversal symmetry with the development of a net magnetization below 2.3~K \cite{kawano1,kawano,Choi,CanfieldII}.
Therefore the pure magnetic states for undoped $R$~=~Er and Ho $R$Ni$_2$B$_2$C are close to each other and more importantly possibly exchangeable.  
The subsititution of Ho sites with Er ions (or vice versa) influences the RKKY exchange interactions and changes the type of order with respect to the pure material. This would also change the crystal electric fields (CEF) and their coupling with the RKKY modulations. These newly obtained states for Ho$_{(x-1)}$Er$_x$Ni$_2$B$_2$C may show new IC and/or C AFM modulations or even destroy some of the original ordering seen in pure systems.

In this study we focus on magnetism, superconductivity and the interplay of these states within Er doped Ho$_{(1-x)}$Er$_x$Ni$_2$B$_2$C single crystals using magnetic susceptibility and triple-axis neutron diffraction techniques. 
    
Single crystals of $R$Ni$_2$B$_2$C; $R$~=~Ho$_{(1-x)}$Er$_x$ with x~=~0, 0.25, 0.50 and 0.75 were grown using a slow-cooling self-flux technique \cite{Cho}. These samples contain isotopically enriched $^{11}$B isotopes to minimize the neutron absorption of natural boron. The magnetization measurements were performed at the Brockhouse Institute for Materials Research (BIMR) at McMaster Univerisity using a Quantum Design Magnetic Properties Measurement System (MPMS). Single crystal samples having flat surfaces for crystallographic $a\textrm{-}b$ planes were aligned so that their c-axis becomes perpendicular to the applied field direction. The neutron diffraction experiments were conducted at NIST Center For Neutron Research (NCNR).  Thermal triple-axis BT-9 and double focusing triple-axis BT-7 instruments \cite{Lynn4} and NG-5 SPINS cold triple-axis neutron spectrometers were used. Both closed cycle and cryogen cryostats were used for performing temperature scans on different instruments. Most of the measurements were performed to the lowest possible temperature (in the vicinity of 1.5 K). The incident neutron energy was set to 14.7 meV for BT-7 and BT-9 and 5 meV for NG-5 measurements. A typical collimation with pyrolytic graphite (PG) filters of 40-40-PG-Sample-PG-40-open was selected for the BT-9 experiments. Measurements on the NG-5 instrument were performed using a Beryllium Filter (BE) with Guide-BE-80-Sample-80-BE-single detector collimation. For the BT-7 measurements a position sensitive detector (PSD) with 80-PG-80R-Sample-PSD collimation was employed. The map of magnetically interesting reciprocal space was collected using the PSD. 
Each PSD data set was collected in 3 different $\theta$ sample angle rotations from 0$^o$ to 100$^o$  where the corresponding $2\theta$ angles were set at 12$^o$,17$^o$ and 22$^o$ for 401 data points resolution. Scans in reciprocal space are presented in the reciprocal lattice units of H~=~$\frac{2\pi}{a}$ and L~=~$\frac{2\pi}{c}$ where a is $\sim$~3.5 \AA ~and c is $\sim$~10.5 \AA.    
Our aim was to study the magnetic structures and the superconductivity of each sample depending on the x amount. 

\begin{figure}
\includegraphics[scale=.9]{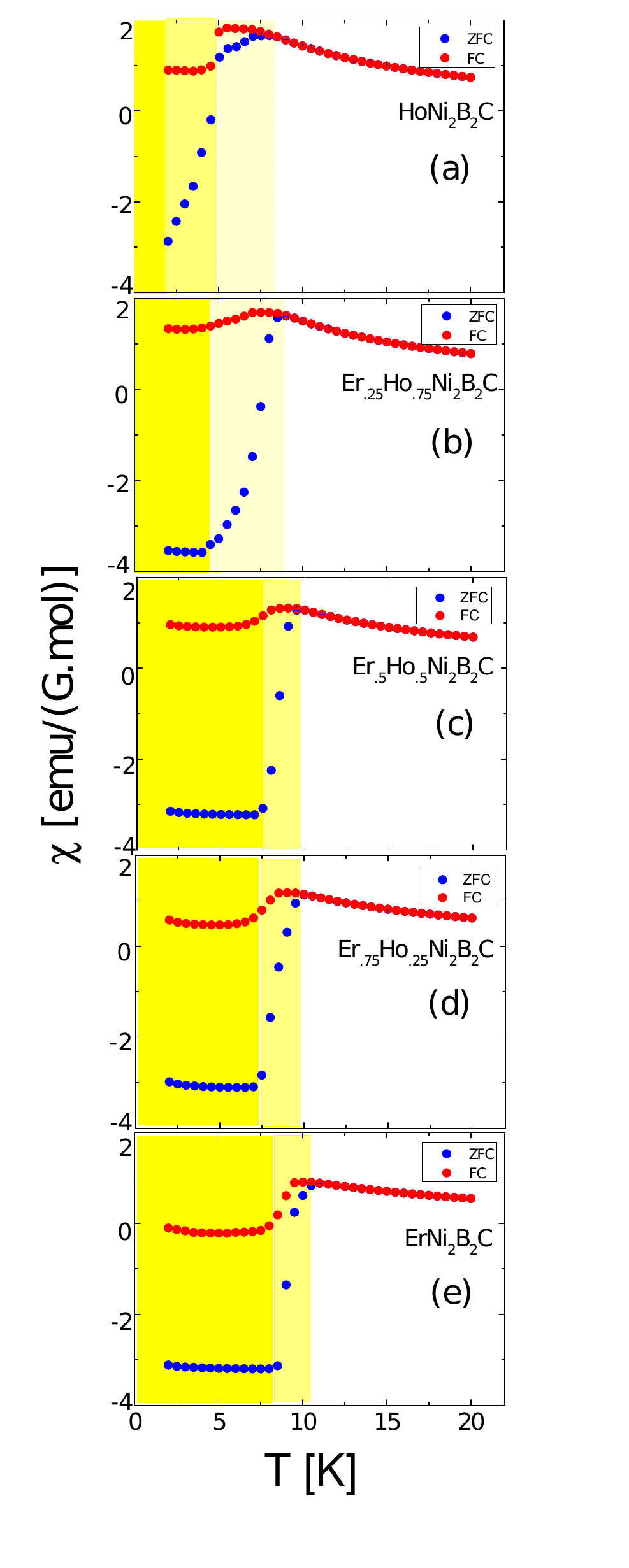}
\vspace{-1cm}
\caption{\textbf{Susceptibility measurements for all samples.}
\footnotesize { The yellow regions mark the superconductivity where diamagnetism is partially or completely observed. Very light yellow regions especially for panel (a) and partially in panel (b) indicate the reentrant superconductivity region in pure HoNi$_2$B$_2$C and Ho$_{1-x}$Er$_x$Ni$_2$B$_2$C where x~=~0.25. One tone darker light yellow regions mark the developing superconductivity.  The offset temperature of Er doped samples marked with dark yellow regions increases from $\sim$ 4.5 K to $\sim$ 9.5 K as x increases from 0.25 to 1 in samples as shown from panel (b) to (e). (Note, 1 emu = 1x10$^{-3}$ Am$^2$)  } 
}
\label{Fig1}
\end{figure}

In order to see the superconducting phase and to determine the transition temperatures as functions of x, we performed a series of DC magnetization measurements, shown in Fig.~\ref{Fig1}. The field cooled (FC) and zero field cooled (ZFC) measurements were conducted for $\sim$~2~$\leq$~T~$\leq$~20 K range. An applied field of 1mT  was used for FC measurements. Each crystal shows regular cleavage sharp corners pointing along one of the principal axis directions on the clean $a\textrm{-}b$ crystallographic plane surfaces. This was confirmed by Laue pictures which are not shown here. Thus we were able to align the field along [100] (or [010]) within $\pm$~5$^o$ accuracy. 

\begin{figure}
\hspace*{-.6cm}
\includegraphics[scale=0.85]{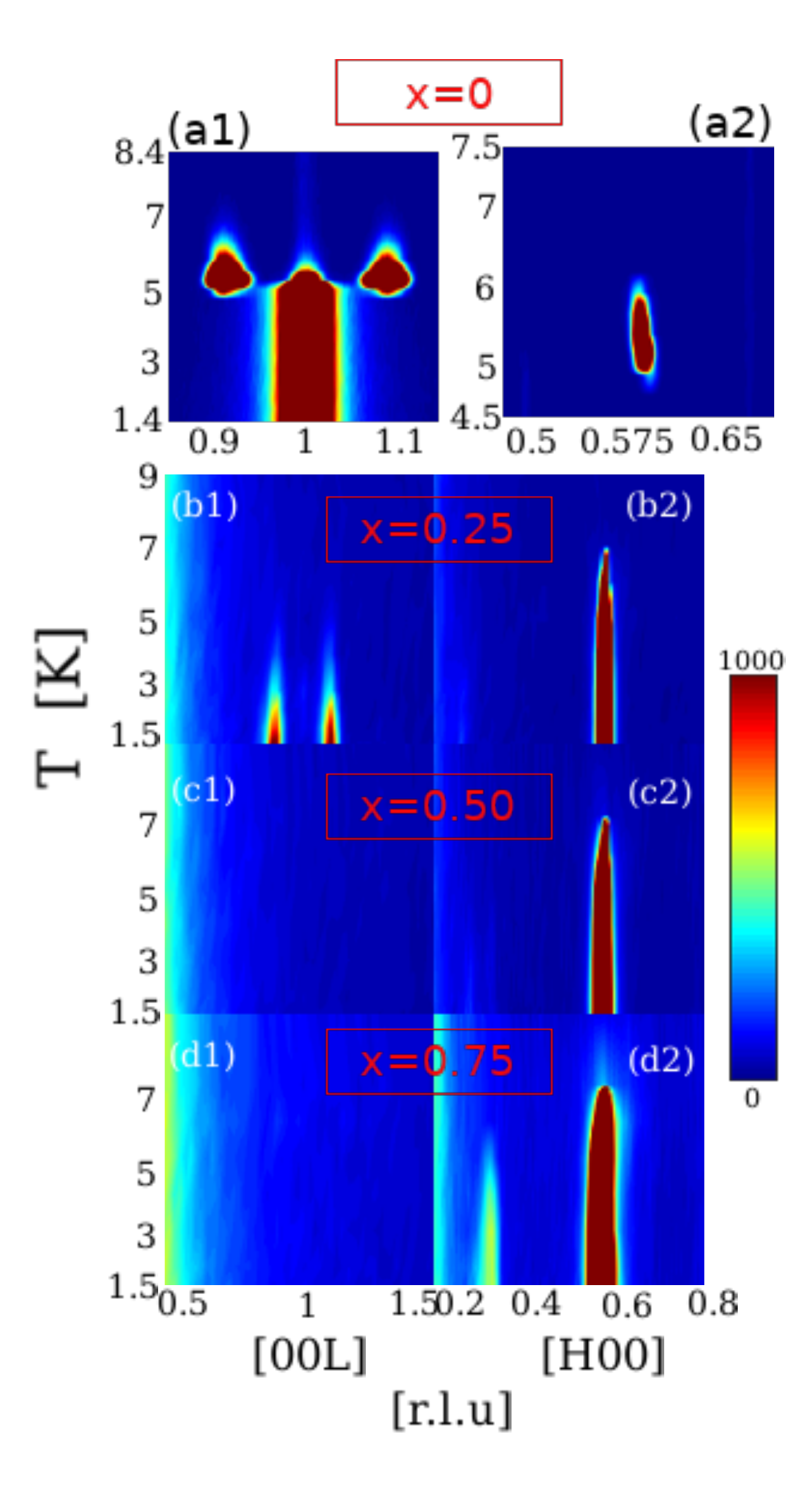}
\caption{\textbf{Systematic temperature study of magnetic reflections along [00L] and [H00] directions.}
\footnotesize{Each neutron diffraction data set shown for panel (b1,b2) to panel(d1,d2) is focused on the interesting region which was set by the pure HoNi$_2$B$_2$C AFM ordering. The pure system x~=~0 Ho$_{(1-x)}$Er$_{x}$Ni$_2$B$_2$C has one commensurate AFM modulation at (001) and two incommensurate satellite AFM modulations with (001$\pm\delta$) where $\delta$=0.09 shown in panel (a1). It has also one incommensurate peak at (0.58 0 0) shown in panel (a2). The same study with a wider region is given for Er containing samples for x~=~0.25 (panel b(1) and b(2)), x~=~0.50 (panel c(1) and c(2)) and x~=~0.75 (panel d(1) and d(2)), in order.  }
}
\label{Fig2}
\end{figure}

The first common property of all data shown in Fig.~\ref{Fig1}. is the paramagnetic behavior over T~$>$~10 K and the net difference in signal between two different cools. 
Below 10 K, the diamagnetic signal of the SC phase becomes dominant. The signal difference between ZFC and FC can be explained by the remnant field trapped in the sample in FC cycles which screens the diamagnetic signal. The FC saturated low temperature susceptibility remains positive for all samples except for pure ErNi$_2$B$_2$C. This is seen in panel (e). There is a net decrease in this value as the Er content increases. However, the positive susceptibility (up to Ho$_{0.25}$Er$_{0.75}$Ni$_2$B$_2$C in panel (d)) is evident for 1mT being enough to penetrate the SC region by partially destroying the diamagnetism. In other words the observed lower critical magnetic field values change as a function of Er concentration and the exact values are in the vicinity of this field value. This is in agreement with the value obtained for pure HoNi$_2$B$_2$C \cite{Memo}.       
In panel (a) the reentrant behavior of the superconducting phase can be seen with a soft decrease in susceptibility at $\sim$ 8 K. This is followed by a sharp decrease at $\sim$ 5 K where the diamagnetic signal is obtained indicating the end of the reentrant region and the start of the SC phase. This also coincides with the disappearing temperature of IC AFM modulations shown in Fig.~\ref{Fig2} (a) for pure HoNi$_2$B$_2$C. The same sudden decrease in signal at the same temperature is also valid for the FC cycle. The reentrant region disappears as the Er content increases in the other samples. This is shown with the 25 $\%$ Er sample shown in panel (b). ZFC data shows a saturated diamagnetism for all Er containing samples. Also, the slow decrease in susceptibility signal seen in pure HoNi$_2$B$_2$C becomes sharper as the Er content increases. Thus while the onset for SC seen in panel (b) is T~$\sim$~8~K for the 25 $\%$ Er sample, it becomes $\sim$ 10 K for pure ErNi$_2$B$_2$C shown in panel(e). The same concept can be seen in the offset temperature values with an increase from T~$\sim$~4.5 K to T~$\sim$~9.5~K for 25 $\%$ Er and for pure ErNi$_2$B$_2$C , respectively. For the Ho$_{0.75}$Er$_{0.25}$Ni$_2$B$_2$C data shown in panel (b), one can suggest the existence of a weak reentrant behaviour for a very small temperature range, however, this entirely disappears for Ho$_{0.5}$Er$_{0.5}$Ni$_2$B$_2$C.

\begin{figure*}
\centering
\vspace{-1cm}
\includegraphics[scale=0.7,trim={0 0.0 0 0},clip]{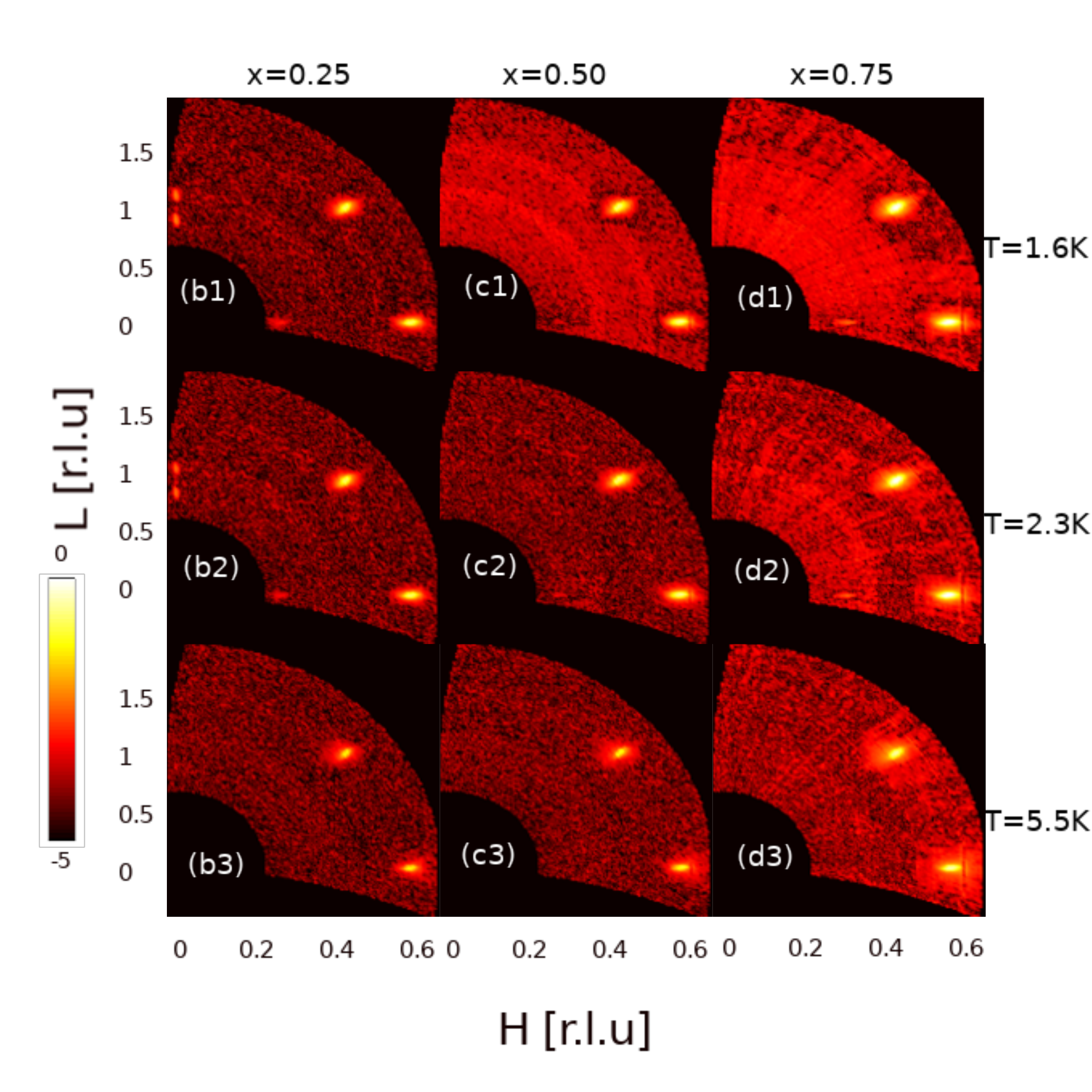}
\caption{\textbf{Magnetic neutron diffraction measurements of scattering plane (H0L).} 
\footnotesize{These data were obtained using a position sensitive detector (PSD) on the BT-7 spectrometer. Three different temperatures shown on the right side of the panels are used in this study. Each panel is built up by concatenating 3 different constant 2$\theta$ scans forming foreground. T~=~9 K data as background for each sample (this is not shown) is subtracted from each foreground data. A logarithmic intensity scale is used, enabling weak reflections to be seen together with strong reflections. For the ease of reading and comparing, the same order for labeling used in Fig.~\ref{Fig2} is continued in this figure.}
}
\label{Fig3}
\end{figure*}

Neutron diffraction measurements along [H00] and [00L] directions centered at H~=~0.5 and L~=~1, respectively, are shown in Fig.~\ref{Fig2}. The main goal for these measurements was to study the Er doping effect as a function of temperature taking the pure  HoNi$_2$B$_2$C as the reference magnetic system. The temperature dependence of the magnetic reflections shown in panel (a1) and (a2) is for the pure HoNi$_2$B$_2$C sample. There are 3 IC peaks which are the characteristic of AFM modulations valid for pure HoNi$_2$B$_2$C shown in Fig.~\ref{Fig2} panel (a1) and (a2). Two of these peaks occur as satellite peaks to the long-range C-AFM order shown in panel (a1). The C-AFM peak on (001) behaves as an order parameter (squared) for pure HoNi$_2$B$_2$C \cite{Grigereit}. The satellite peaks in this direction start to order at almost the same temperature as with the IC modulation seen on (0.58 0 0) in panel (a2). These 3 IC - AFM reflections are observed between 6 K to 5 K. At T~$\sim$~5 K the C-AFM peak which is shown in panel (a1) on (001) starts to order while at this temperature the IC-modulations start to dissolve altogether. The commensurate magnetic modulation intensity shows a discontinuous order. 
The temperature maps shown in these two panels serve as the reference for magnetic behaviour of other samples which contain some amount of Er. When the Er content is just x~=~0.25 in the Ho$_{(1-x)}$Er$_{x}$Ni$_2$B$_2$C the C-AFM phase is totally destroyed and IC magnetic modulations along L direction order at $\sim$ 4 K, shown in panel (b1). This temperature is lower than the ordering temperature of pure HoNi$_2$B$_2$C. Surprisingly, the wave-vectors for these IC modulations are essentially the same as the parent compound, while the intensities keep increasing with decreasing to the base temperature. The IC modulation seen on (0.58 0 0) for pure HoNi$_2$B$_2$C  still exists for x~=~0.25, however, this composition behaves as a prominent order peak describing an IC-AFM order as shown in panel (b2). 
In panel (c1) and (c2) we see the doping effect of more Er which changes the magnetic behaviour even further away from the reference. The most crowded and thus most interesting scattering plane shown in panel (a1) for the pure system is now empty, without any reflection for x~=~0.5. This means that all magnetic moments in the x~=~0.5 sample aligned themselves with a modulation along the H-direction, resembling the pure ErNi$_2$B$_2$C system. So the spin-density wave (SDW) modulation originally belonging to pure ErNi$_2$B$_2$C becomes more pronounced than the AFM orders seen in pure HoNi$_2$B$_2$C. This behaviour continues for the x~=~0.75 sample and this is shown in panel (d1) and (d2). Different than all other samples, for the x~=~0.75 sample there is another IC magnetic development at the (0.32 0 0) position. This peak's intensity increases at low temperatures shown in panel (d2) and the details are given below. Thinking that the origin of this new type of  reflection might be a higher order satellite, specifically 3$^{rd}$ harmonic modulation peak seen in pure ErNi$_2$B$_2$C, we performed position sensitive diffraction (PSD) measurements in the [H0L] magnetic scattering plane to find all possible magnetic Bragg peaks in this plane.              

\begin{figure*}
\vspace*{-4cm}
\includegraphics[scale=0.7,trim={0cm 3cm 1cm 3cm}, clip]{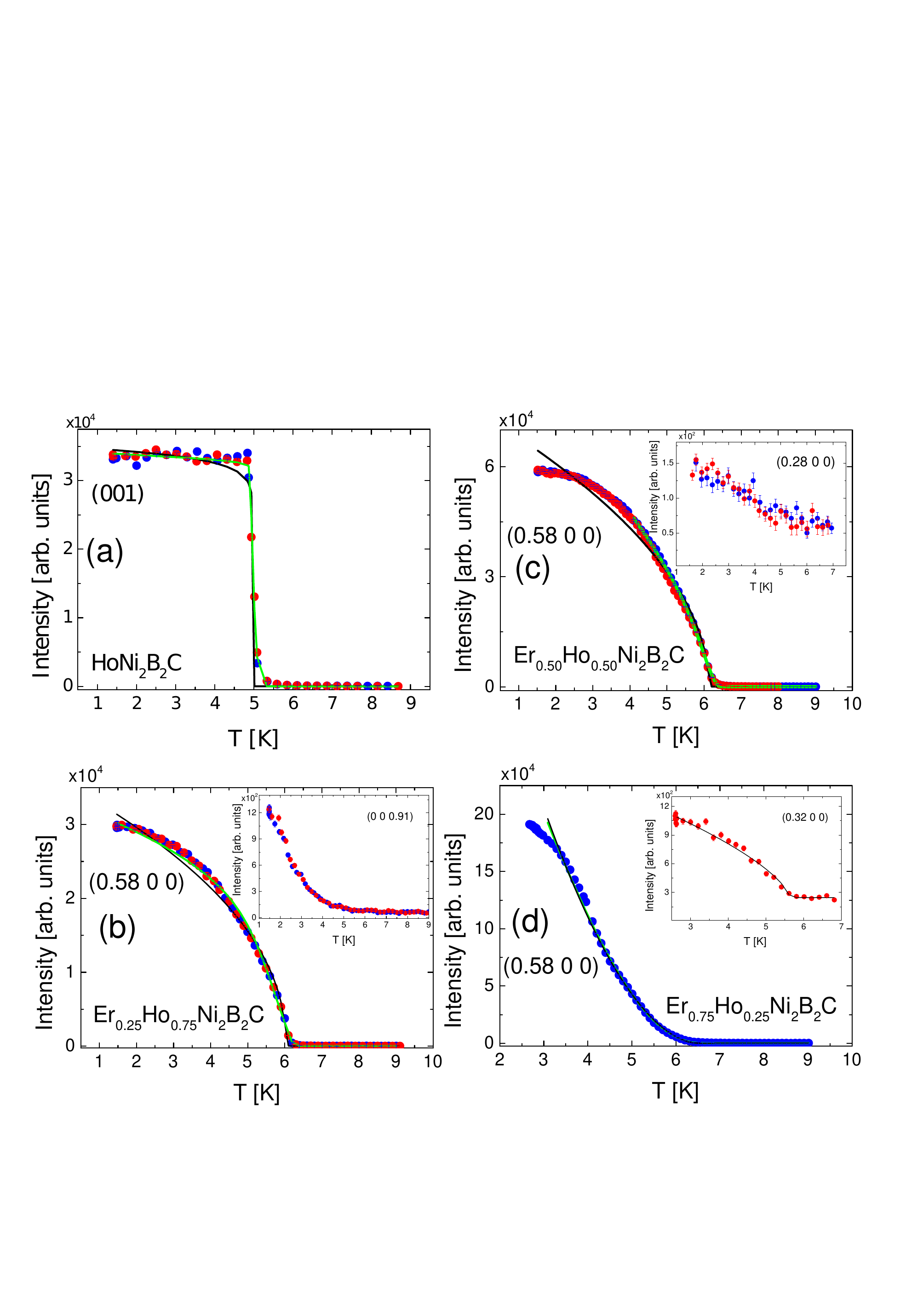}
\caption{\textbf{Order parameter fit analyses for different magnetic peaks for x~=~0, 0.25, 0.5 and 0.75 samples.}
\footnotesize{The fit results are given in Table.~\ref{tablo} }
}
\label{Fig4}
\end{figure*}

\begin{table}
\caption {
\textbf{Order parameter fit results.}
\footnotesize{The results of two power law fits shown by black and green lines in Fig.~\ref{Fig4} are given. The units of N\'eel  order temperature T$_N$, the superconducting critical temperature T$_C$ and the FWHM ($\Gamma$) are Kelvin. }
\vspace{1mm}
}

\label{tablo}  
\hspace*{-.5cm}
\begin{tabular}{|c|c|c|c|c|}
\hline
 
Black&~x=0~          &~x=0.25~         &~x=0.50~         &~x=0.75 ~\\
Line&$\beta$=0.009(10)&$\beta$=0.15(1)&$\beta$=0.36(1)&$\beta$=0.78(30)\\  
~     ~&~T$_N$=4.9(1)~    &~T$_N$=6.20(1)  ~&~T$_N$=6.24(1) ~ &~T$_N$=6.7(5) ~\\
\hline
Green&~$\beta$=0.02(1)~&~$\beta$=0.155(8)~ &~$\beta$=0.313(3)~&~For inset fit~\\  
Line&~T$_N$=4.97(5)~    &~T$_N$=6.27(2)  ~&~T$_N$=6.42(2) ~  &~$\beta$=0.33(15) ~\\
~     ~&~$\Gamma$=0.58(2)~&~$\Gamma$=0.69(7)~&$\Gamma$=0.35(7)~ &~T$_N$=5.63(5)~\\
\hline
       &~T$_C\sim$5K~     &~T$_C\sim$7.5K~ &~T$_C\sim$8K~& ~T$_C\sim$8.5K~     \\        
\hline

\end{tabular}\\ 
\vspace{1mm}
\end{table}

The magnetic modulations studied in temperature scans in the previous figure, are now shown in Fig.~\ref{Fig3} as diffraction from the [H0L] scattering plane for 3 different temperatures. These measurements were conducted using a PSD on the BT-7 instrument for 3 different Er containing samples, x~=~0.25, 0.50 and 0.75. For each study, the magnetically important region which is mostly the first Brillouin zone and some part of second one, forming a foreground, was investigated at T~=~1.6 K, 2.3 K and 5.5 K. The T~=~9 K data were used for each sample as background (not shown) and subtracted from the data. A logarithmic intensity scale is used to reveal the weak intensity peaks. 
Therefore the data shown in each panel in Fig.~\ref{Fig3} contain solely magnetic intensities. These 3 temperatures were chosen in order to show a more inclusive view of scattering both in the ordered phases and during transitions. 1.6 and 2.3 K data are used to determine the effect of weak FM order seen in pure ErNi$_2$B$_2$C (x~=~1) samples while the 5.5 K data are important to isolate the IC ordering especially valid for pure samples.
The satellite peaks at (0 0 0.91) and (0 0 1.09) indicating IC-AFM modulations were seen for the x~=~0.25 sample (panel (b1)) at T~=~1.6 K while they don't exist for other Er dopings at this temperature. As temperature increases towards T~=~5.5 K these peaks lose their intensities and disappear. This is shown in panel (b2) at T~=~2.3 K and in panel (b3) at T~=~5.5 K . Interestingly, there is an IC peak appearing at (0.28 0 0) best observed for x~=~0.25 shown in panels (b1) and (b2). This peak appears for the x~=~0.50 sample at T~=~1.6 K and 2.3 K but the intensity is weak compared to the ones for x~=~0.25. The dominant peaks for all samples occur at (0.58 0 0) and (0.58 0 1) arising from the scattering in the 1$^{st}$ and 2$^{nd}$ Brillouin zones. These two peaks exist at all temperatures, shown in Fig.~\ref{Fig3}. 

The magnetic intensities shown in Fig.~\ref{Fig2} and Fig.~\ref{Fig3} have been studied as a function of temperature with finer data sampling in both cooling and heating directions. For this study the magnetic reflections having order parameter like  profiles were selected. The results are shown in Fig.~\ref{Fig4}. In panel (a) of this figure, the (001) magnetic reflection which is the C-AFM modulation wave-vector for the pure HoNi$_2$B$_2$C is shown. The phase transition occurs at T~$\sim$~5~K with a 1$^{st}$ order phase transition character. This is evident from the sudden increase in the observed intensity. The black line in this figure represents the (square of the) order parameter power law fits which are as 

\begin{equation}
I(T) \sim (1-\frac{T}{T_N})^{2\beta}
\label{equ1}
\end{equation}

\begin{equation}
I(T) \sim \int G(T_N,\Gamma)(1-\frac{T}{T_N})^{2\beta}dT
\label{equ2}
\end{equation}

where t~=~$\frac{T}{T_N}$ is the reduced temperature. A more complicated power law fit including a Gaussian (G(T$_N$,$\Gamma$)) to present a distribution of transition temperatures immersed at the center of the transition temperature  at T~=~$T_N$ was also used in the analyses. This is given in Equ.~\ref{equ2} and the result of its fit is shown in the green colored lines in Fig.~\ref{Fig4}. The calculated full-width-at-half-maximum (FWHM) obtained from the Gaussian line can be used as another estimate for the strength of the broadening of the phase transition as well as with $\beta$. Equ.~\ref{equ2} is needed to  capture the smearing effects during transition which are possibly created by the existing disorder sources \cite{simon,memo2}. In our samples two rare-earth magnetic elements of Er and Ho would have different CEF which act on one another to create some sort of disorder effects. Table ~\ref{tablo} shows the results of the fits for these analyses. As seen from the panels of Fig.~\ref{Fig4} and the corresponding values given in Table~\ref{tablo}, the order parameter exponent $\beta$ increases as the Er concentration increases. The IC peaks lying on the (0.58 0 0) magnetic reflection exhibit an order parameter for the two samples of x~=~0.25 and 0.50. However, the peak profile for x~=~0.75 differs slightly from this, shown in panel (d). The fitted $\beta$ value becomes larger than  the mean field value 0.5 for this reflection. At this point another magnetic reflection also catches some attention. The IC modulation which is the satellite peak of pure HoNi$_2$B$_2$C at (0 0 0.91) does not disappear below T~=~5 K like it does in the pure system. This was shown previously in Fig.~\ref{Fig2}. The temperature behaviour of this peak is now shown in the inset of panel (b) in Fig.~\ref{Fig4}. This I(T) profile can be interpreted as induced magnetism \cite{Lynn3}. The same argument is also valid for the peak profile of the x~=~0.75 sample. In addition, the peak profile shown in the inset of panel (c) suggests a growing order like behavior for the (0.28 0 0) IC satellite. Finally, the (0.32 0 0) peak profile shown in the inset of panel (d) can be fit to a power law and the results are given in Table ~\ref{tablo}.

The critical order parameter analysis results given in Table I exhibits several outcomes. The pure HoNi$_2$B$_2$C  long range AFM order obtained from the magnetic [100] reflection is a 1$^{st}$ order phase transition. This has been previously obtained and can also be visibly confirmed with our results , Fig.~\ref{Fig4} panel (a) \cite{Kressig, Gupta}. With our analysis, the power law exponent $\beta$ is fit to nominally zero which is the hallmark of a discontinuity. Therefore, the  discontinuous character of a 1$^{st}$ order phase transition is numerically observed. 
As the Er amount increases the RKKY exchange interactions between Er and Ho ions including a possible local frustration on each other's magnetic modulations drive the order character towards the 3D-XY exponent values seen for the x~=~0.50 results. This argument is fortified by values obtained for the x~=~0.25 sample where the obtained $\beta$ value is in the vicinity of those documented for 2D-XY frustrated systems \cite{Landau,Darton}. When x~=~0.75 is considered, the magnetic order observed on the (0.58 0 0) magnetic reflection is no longer a conventional order parameter profile. Instead it suggests an induced magnetic order with possible fluctuations where the intensity profile resembles the induced $R$  moments under the influence of Cu order seen in $R_2$CuO$_4$, $R$~=~Nd and Pr for example \cite{Lynn3}. The induced magnetic order argument is also supported by the two other magnetic reflections shown in the insets of panel (b) and (c) in Fig.~\ref{Fig4} for x~=~0.25 and 0.50, respectively.
We remark that this clear induced order implies strong coupling between the two order parameters, which in the case of cuprates mentioned above occurs because the orderings have the same symmetry. This contrasts sharply with Sm$_2$CuO$_4$, where the Cu and Sm order parameters are different and no significant coupling is observed \cite{Sumarlin}.   
There are also IC magnetic reflections. In addition, the frustration between the Ho and Er moments can yield new IC magnetic orders such as the one shown with a power law fit in the inset in panel (d) of Fig.~\ref{Fig4}.     
We see that doping Er into HoNi$_2$B$_2$C with 25 $\%$ which is the smallest amount in this study, entirely suppresses the commensurate AFM long range order in the pure parent compound. Doping with Er also changes the IC-AFM temperature dependence. When the magnetization and susceptibility results are compared, we can conclude that superconductivity and the IC magnetic orders do not strongly compete with each other. Instead they tend to order at low temperatures. With x~=~0.50 and higher values the overall magnetic intensity profiles start to look like the ones observed for pure ErNi$_2$B$_2$C with no magnetic reflection along the L-direction and with only IC intensities in the H-direction. Finally, for x~=~0.75 a power law behavior becomes insufficient to capture the overall intensity temperature profile shown in panel (d) of Fig.~\ref{Fig4}. The peak profiles shown for those IC magnetic reflections clearly suggest the existence of dominant induced magnetic character similar to the ones  previously studied in pure and doped $R_2$CuO$_4$, $R$~=~Nd and Pr \cite{Lynn3}. 

In this study Er doped Ho$_{1-x}$Er$_x$Ni$_2$B$_2$C single crystal samples with x~=~0.25, 0.50 and 0.75 were investigated using MPMS and triple-axis neutron diffraction techniques. Further studies using small angle neutron scattering to investigate the effect of the magnetic order on the Abrikosov lattice would be an interesting new avenue to pursue.

$Acknowledgement$
M.R. and S.G. were supported by TUBITAK grant no. 1001-117F315. The identification of any commercial product or trade name does not imply endorsement or recommendation by the National Institute of Standards and Technology, NIST.

\
\bibliography{ErHoNiBC_Bibliography_List}

\end{document}